\documentclass[%
 reprint,
 amsmath,amssymb,
 aps,
showkeys,
]{revtex4-2}

\usepackage{graphicx}
\usepackage{dcolumn}
\usepackage{bm}

\newcommand{\HZ}{\text{Hz}}

\begin{document}

\preprint{arXiv: April 1, 2026}

\title{Mexican Burrowing Toads as gravitational wave detectors}

\author{Frederic V. Hessman}
\affiliation{%
 Institut~f\"ur~Astrophysik und Geophysik, University~of~G\"ottingen
}%

\author{Christian Jooss}
\affiliation{%
 Institut~f\"ur~Materialphysik, University~of~G\"ottingen \\
}%



\bigskip

\date{April 1, 2026} 

\begin{abstract}
It is generally assumed that gravitational waves are extremely difficult to detect. 
However, we show that the call of the Mexican Burrowing Toad has an amazing resemblance to cosmic gravitational wave signals due to the merging of neutron stars and/or black holes.
It is known that toads exhibit magnetoreception -- the ability to detect magnetic fields -- and that magnetic fields thus subtly affect ion channel activities in toad neurons. 
We speculate that gravitational strains produce phonons and magnons in a ferromagnetic substance embedded in the nervous system of the toads and that these coherent signals are exponentially amplified by a Raman laser mechanism to the point where they can be detected.
The fine tuning necessary for this mechanism to work would help to explain why this species of toad show this remarkable ability and others do not.
We analyze the sound of a pond full of Mexican Burrowing Toads in the hopes of detecting slight phase shifts in their calls due to a gravitational wave event.
No effect was found and the the LIGO/VIRGO consortia have not reported an event during the recording, illustrating the power of this approach.
We suggest the massive use of these toads would be an inexpensive way to support the operation of optical interferometric gravitational wave detector facilities.
\end{abstract}

\maketitle

\begin{figure}[!b]
    \centering
    \includegraphics[width=8cm]{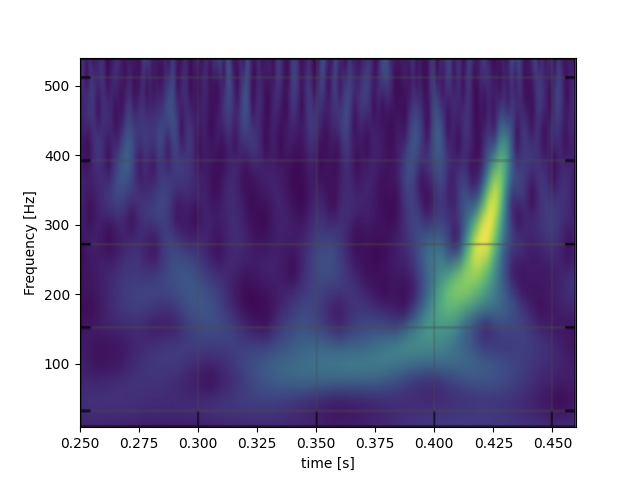}
    \caption{Time-resolved spectrogram of the strain amplitude for the GW event GW150914 \cite{Abbott2016}.  The chirp frequency goes from approximately 100 to 400\,Hz within about 100\,ms.}
    \label{fig:GW150914}
\end{figure}

\section{Introduction}

The detection of gravitational waves (GW) marked a transformative moment in physics, confirming a key prediction of Einstein's General Theory of Relativity \citep{Einstein1916}. These ripples in spacetime, first observed by the Laser Interferometer Gravitational-Wave Observatory (LIGO) in 2015 \citep{Abbott2016}, were generated by the merger of two black holes (BH) \citep{SHC}. The event, designated GW150914 (Fig.\,\ref{fig:GW150914}), provided direct evidence of binary black hole coalescence and demonstrated the immense power released during such cosmic collisions.

A distinctive feature of GW signals from binary mergers is their characteristic "chirp" pattern, where both the frequency and amplitude increase as the two massive objects spiral toward each other. This waveform results from the loss of orbital energy due to GW emission, culminating in the final inspiral, merger, and ringdown phases \cite{Maggiore2008}. The mathematical modeling of such waveforms is crucial for interpreting observational data and understanding the properties of the merging objects.

\begin{figure}[b!]
    \centering
    \includegraphics[width=8.6cm]{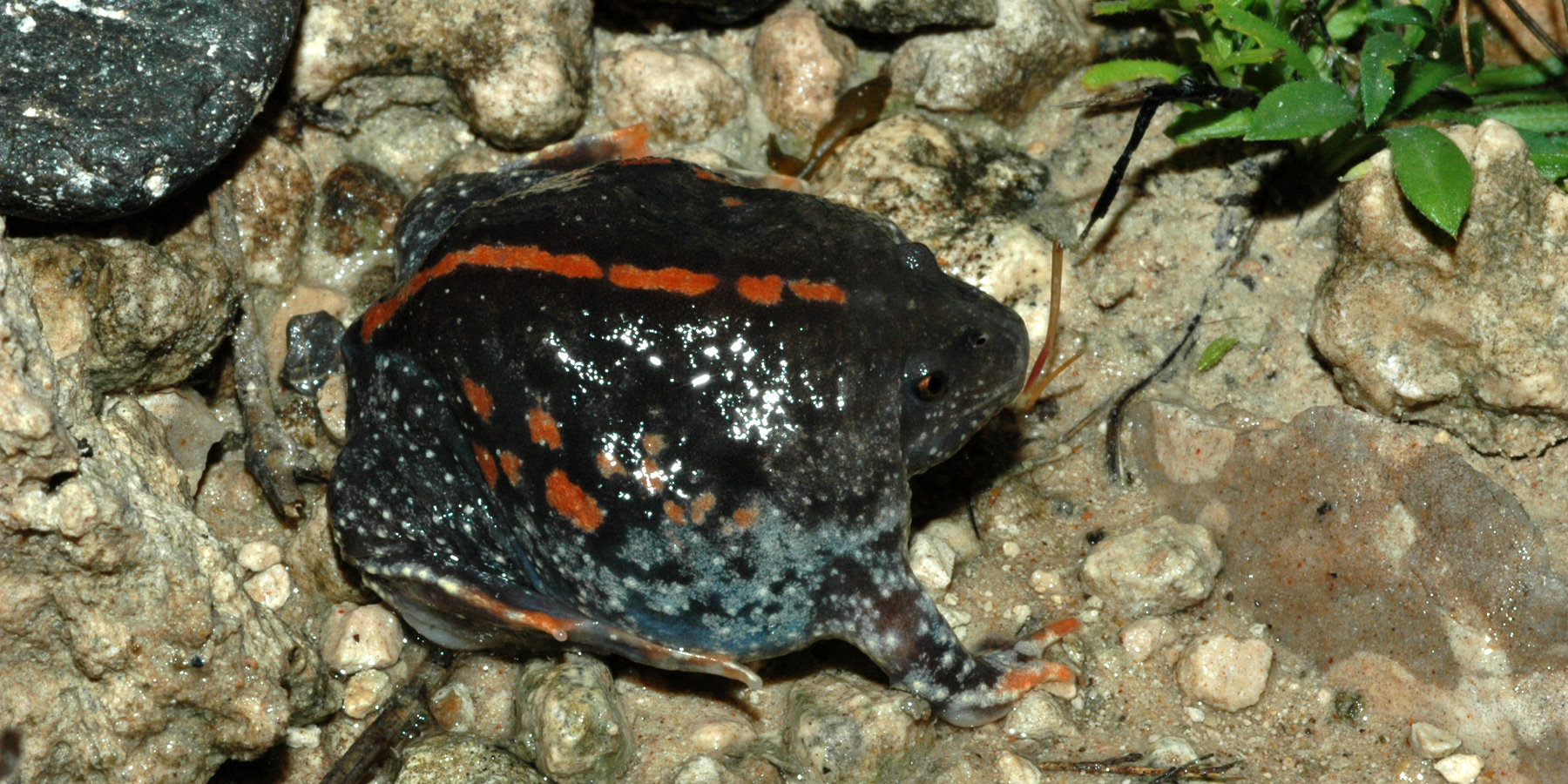}
    \caption{Image of a Mexican Burrowing Toad ({\it Rhinophrynus dorsalis}), courtesy of \cite{toad}.}
    \label{fig:toad}
\end{figure}

Interestingly, the chirp-like nature of these GW bears a striking resemblance to the call of the Mexican Burrowing Toad ({\it Rhinophrynus dorsalis}, hereafter MBT; see Fig.\,\ref{fig:toad}). This amphibian produces a distinct call during mating, which, like GW signals, starts at a lower frequency and rapidly increases in pitch \cite{Ryan1985}. Though arising from vastly different physical processes, the waveform similarity is remarkable enough to suggest further study.

\begin{figure*}
    \centering
    \includegraphics[width=16cm]{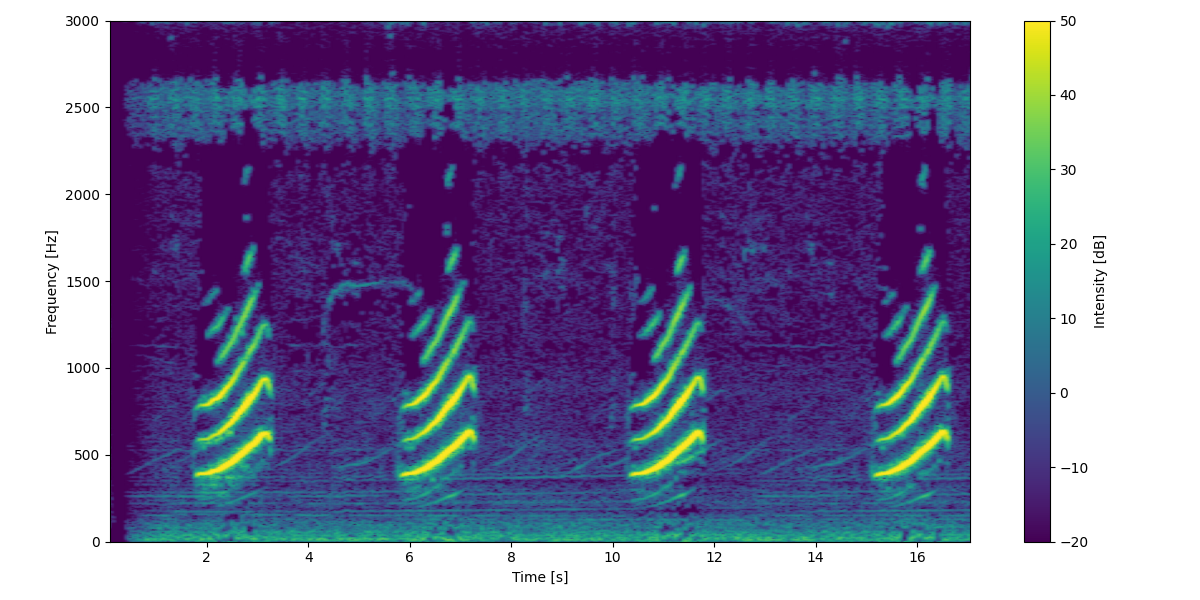}
    \caption{Sonogramme of multiple calls by a single Mexican Burrowing Toad.  Note the rapid rise in frequency displayed by all of the harmonics. A final slight drop in frequency is also visible. In the background, calls by other toads can be seen, some at slightly different frequencies.  The oscillating band at higher frequencies is due to crickets.}
    \label{fig:sonogram}
\end{figure*}

\begin{figure*}
    \centering
    \includegraphics[width=16cm]{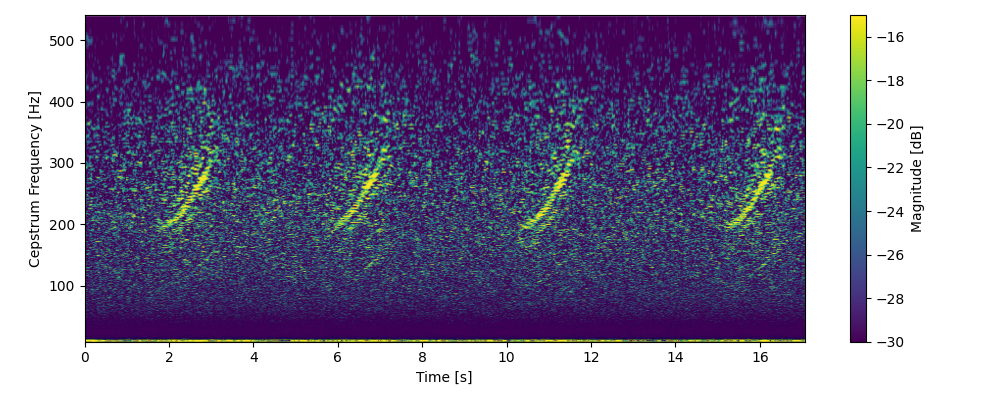}
    \caption{Cepstrogramme of the Mexican Burrowing Toad calls shown in Fig.\ref{fig:sonogram}.  Note how a single signal with a rapid rise in frequency from about 190 to 325\,Hz within about 1250\,msec is seen.   Compare this form with that of the gravitational event shown in Fig.\,\ref{fig:GW150914}.}
    \label{fig:cepstrogram}
\end{figure*}

\section{Analysis of the MBT croak}

A recording of a single MBT as well as that of a whole pond full of them is available in the form of an MP3 audio file at the ``Frogs and Toads of North America'' web-site \footnote{https://archive.org/details/FrogsAndToadsOfNorthAmerica}.
We extracted four isolated calls of a MBT from the original recording using the public-domain audio programme {\tt Audacity} \footnote{https://www.audacityteam.org} (version 3.7.3).
Then, we mixed the stereo audio to mono.
Thereafter, python scripts were used to read the {\tt *.WAV} file exported by {\tt Audacity} and to analyze the signal.

\subsection{Sonogramme}

We first produced a sonogramme of the data -- an image showing the contained frequencies in the audio signal from the toad and the background pond versus time (Fig.\,\ref{fig:sonogram}).

The sonogramme shows several interesting features: the chirp of the MBT's call is clearly seen, mainly as a series of harmonics.
The spacing of the harmonics indicates that the 1st strong component is not the fundamental but the 1st harmonic; a very faint sign of the fundamental can be seen.
Also visible is a slight drop in frequency at the end of the call.
The oscilating band above the call is due to crickets and the general drop in background power during the MBT call is due to the induced cessation of environmental noise.

\subsection{Cepstrogramme}

In order to better isolate the ``chirp'' of the toad's call, we also constructed a cepstrogramme -- basically, the Fourier transforms of the individual spectra in the sonogramme -- that shows the frequencies of signals with high harmonic content as a function of time.
The high-frequency noise was filtered out with a Gaussian low-pass filter of width 400\,Hz.
Finally, the median background cepstrum was subtracted.

The resulting cepstogramme (Fig.\,\ref{fig:cepstrogram}) shows very clean chirp signals with a rapid rise in frequency from about 190 to 325\,Hz within about 1250\,msec.
Compare this with the GW signal in Fig.\,\ref{fig:GW150914}, where the chirp frequency goes from approximately 50 to 400\,Hz within about 100\,ms.
Thus, the frequency bands in both signals occur in the same range, only the lengths of the signals are different.

The chirp frequencies are fairly stable for a single MBT and the background of the sonogramme and cepstrogramme shows fainter calls by other MBT at only slightly different frequencies.

\subsection{Fourier analysis}

It would not be too surprising if MBT were not able to faithfully mimic cosmic BH coalescences with their croaks, but the similar frequence range is quite striking.
What if the chirp is faithfully rendered -- could we learn something about cosmic events from a toad?

We adopt the semi-analytic waveform of non-precessing coalescing binary black holes derived by \cite{2007CQGra..24S.689A}.
The phenomenological Fourier waveform for a coalescing BH binary of total mass $M$ and spin parameter $\chi$ is a function of 5 parameters --
three typical frequencies $(f_{merge},f_{ring},f_{cut})$, a dispersion parameter $\sigma$, and the spin-parameter
\begin{equation}
    \chi \equiv \frac{1}{2} \left( (1+\delta) \frac{S_1}{m_1^2} + (1-\delta) \frac{S_2}{m_2^2} \right) ~ \approx ~ {\cal{O}}(1)
\end{equation}
where $m_i$ are the BH masses, $S_i$ are the spin angular momenta, and $\delta \equiv (m_1-m_2)/M$.

The strain in Fourier space is
\begin{equation}
    h(f) = A(f) e^{-i \Psi(f)}
\end{equation}
where
\begin{equation}
    A(f)  =  C f_1^{-\frac{7}{6}} \left\{
    \begin{array}{ll}
        \left(\frac{f}{f_1}\right)^{-\frac{7}{6}} \left(1\!+\!\sum_{i=2}^3 \alpha_i v^i\right), &  
        f\!<\!f_1 \\
        w_m \left(\frac{f}{f_1}\right)^{-\frac{2}{3}} \left(1\!+\!\sum_{i=1}^{2} \epsilon_i v^i\right), & 
        f_2\!<\!f\!<\!f_3\\
        w_r {\cal{L}}(f,f_2,\sigma), &
        f_1\!<\!f\!<\!f_2, \\
    \end{array} \right.
\end{equation}
is the amplitude and
\begin{equation}
\Psi(f) = 2\pi f t_0+ \phi_0 + \frac{3}{128 \eta v^5} (1+\sum_{k=2}^7 v^k \psi_k)
\end{equation}
is the phase;
$C$ is a constant,
$t_0,\phi_0$ are the time and phase of arrival,
$v \equiv (\pi M f)^{1/3}$,
$\mu_i = \left[f_{merge},f_{ring},f_{cut},\sigma\right]$,
\begin{equation}
\epsilon_1 \! \approx \! 1.4547 \chi \!-\! 1.8897,
~~~
\epsilon_2 \! \approx \! -1.8153 \chi \!+\! 1.6557
\end{equation}
are fits to numerical waveforms,
${\cal{L}}(f,f_2,\sigma)$ is a Lorentzian function centered at $f_2$ with width $\sigma$,
$\omega_m$ and $\omega_r$ are factors that insure the signal is continuous across the three bands,
and the $\alpha_i$ are post-Newtonian corrections for the $l\!=\!2$, $m\!=\!\pm 2$ waveforms:
\begin{equation}
\alpha_2 \! = \! -\frac{323}{224} \!+\! \frac{451 \eta}{168},
~~~
\alpha_3 \! = \! \frac{1}{\chi} (\frac{27}{8} \!-\! \frac{11 \eta}{6})
\end{equation}
where $\eta = m_1 m_2 / M^2$
and the model parameters 
$\psi_k(\eta,\chi)$ and 
$\mu_k(\eta,\chi)$
are tabulated by \cite{2007CQGra..24S.689A}.

We could attempt to fit the observed MBT signal with this functional form by calculating the centroids of the primary ceptogrammes in the main chirp, and by fitting the above parameters with a MCMC fitter \citep{2013PASP..125..306F}, but this function is very cumbersome and it is not clear whether it would be worth the effort.
Thus, we leave this exercise to future studies and attempt in the next section to derive rough estimates only.

\subsection{Interpretation}

When the MBT chirp has a frequency of about 250\,Hz, the time-derivative is about  100\,Hz/s and the signal disappears at about 300\,Hz.
From these numbers, we can roughly calculate the physical properties of the associated GW event.

The detector-frame chirp-mass (i.e.~uncorrected for the effects of cosmological redshift), is
\begin{eqnarray}
    {\cal{M}}_{det} & \equiv & 
    \frac{q}{(1+q)^2} M_{det}
    ~ \approx ~
    \frac{c^3}{G} \left[ \frac{5 \dot{f} f^{-11/3}}{96 \pi^{8/3}} \right]^{3/5} \\ \nonumber
    & \approx & 0.46\,M_\odot 
    \left(\frac{\dot{f}}{100\,\HZ/s}\right)^{3/5}
    \left(\frac{f}{250\,\HZ}\right)^{-11/5}
\end{eqnarray}
(e.g. \citep{blanchet}), where $M_{det}$ is the total detector-frame mass and $q$ is the mass-ratio of the binary.
The latter can be estimated using the merger frequency $f_{merge}$:
\begin{equation}
    M_{det} \approx \frac{c^3}{6^{3/2} \pi G f_{merge}} ~ = ~
    15\,M_\odot \left(\frac{f_{merge}}{300\,\HZ}\right)^{-1}
\end{equation}
which implies a mass-ratio
$q \approx 0.003$ and a secondary mass $M_2$ of only $0.05\,M_\odot$, i.e. about the mass of a brown dwarf (BD).
Correcting for cosmological redshift wouldn't change the mass-ratio but would make the masses even smaller.

What secondary objects are possible for this scenario?
They would fill their Roche lobes with mean radius $R_R(q)$ when their mean density is 
\begin{equation}
    \overline{\rho_R(q)} \, \approx \frac{3\pi}{G P_{orb}^2} \frac{q}{1+q} \frac{1}{R_R(q)^3}
\end{equation}
Since $f_{GW} \!=\! 2 f_{orb} \!\equiv\! 2/P_{orb}$ (quadrapole symmetry), and $R_R(q)/a \!\approx\! 0.49 q^{1/3}$ for $q << 1$, the mean density of the secondary must be
\begin{equation}
    \overline{\rho_2} \, \approx 3\times 10^{16}\,kg/m^3 \left(\frac{f_{GW}}{300\,Hz}\right)^2 
\end{equation}
i.e. ruling out a material object like a  BD, white dwarf or a neutron star; only a BH secondary with a BD mass is consistent with these numbers.
There is no known stellar evolutionary path to such low-mass BH, so these objects would have to be primordial, presumedly the source of Dark Matter.
However, this solution has been ruled out by the phenomenon of {\it Spontaneous Human Combustion} \citep{SHC}. 

Thus, we must conclude that, while toads may be capable of detecting GW, they are not good at exactly mimicing them.

\begin{figure*}
    \centering
    \includegraphics[width=18cm]{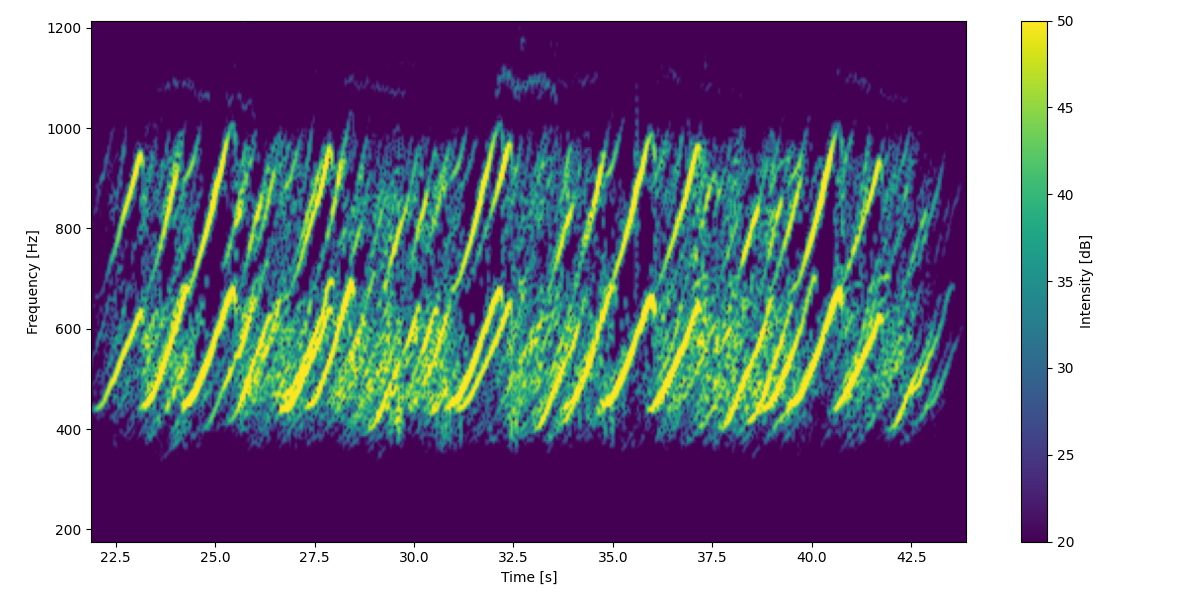}
    \caption{Time-resolved sonogramme from a recording of a pond full of Mexican Burrowing Toads.}
    \label{fig:pond}
\end{figure*}

\section{The biophysical detection mechanism}

The technical effort needed to detect GW using laser interferometers is immense \citep{GW}, so the ability of toads to do the same with biological means is astounding.
On the other hand, the ability of some animals to detect physical processes like slight variations in magnetic fields is also astounding and occurs in a wide range of reptiles and birds \citep{MouritsenRitz2005}.

The mechanisms in biological magnetoreception are varied and still not well-understood \citep{KirschvinkWalkerDiebel2001, JohnsenLohmann2005}.
There are a host of possible magnetic and electric couplings possible \citep{HoreMouritsen2016}, including non-synaptic coupling \citep{Chang2018} at very low levels of electric fields (less than $\sim mV/mm$).
While signals can be processed at roughly thermal energies ($\sim 0.1 eV$), the strains induded by GW are so small (about $10^{-17}\,\mu \epsilon$) that some massive amplifying mechanism is needed.

We can attempt to reverse-engineer this mechanism.
Assume the GW event produces a local oscillating strain in a ferromagnetic material.
Such a very coherent strain -- while very small -- is qualitatively different from the usual acoustic strains due to thermal noise and hence could result in the coherent production of phonons and magnons throughout the material.
It is well known that the interactions between phonons with magnons can result in the transfer of energy between both \citep{magnon-phonon}.
If there is some form of coherent resonance, this can result in the amplification of the magnons, possibly to the point that these oscillations are detectable by a toad's nervous system.
What we need is a magnetronic Raman laser mechanism.

We have constructed a simple resonance model for the response of a ferromagnetic solid to the tiny but coherent strains caused by GW \cite{raman}.
The modified Landau-Lifshitz equation for the magnetisation of the material is then 
\begin{equation}
    \frac{d\vec{M}}{dt} \approx
        -\gamma \vec{M}\!\times\!\vec{H}
        - \frac{\lambda \vec{M}\!\times\!(\vec{M}\!\times\!\vec{H})}{M^2} 
        + \frac{\alpha_{c} \omega_{GW}^3 \vec{M}}{(\omega_{beat}\!-\!\omega_{GW})^2}
\end{equation}
where $\omega_{GW}$ is the strain frequency, $\vec{M}$ is the magnetisation, $\vec{H}$ is the local seed magnetic field, $\gamma$ is the material's gyromagnetic ratio, $\lambda$ is a damping constant,
$\omega_{beat}$ is the beat frequency between the induced phonon and magnon waves, and $\alpha_{c}$ is the magnon-phonon resonant coupling parameter.
Clearly, if the strain frequency approaches the phonon-magnon beat frequency, an exponential magnification of the magnetisation occurs.
%

Apparently, MBT have ferromagnetic materials that result in a large magnification, whereas other toads with other materials probably have different $\omega_{beat}$ and/or $\alpha_c$ where there is no exponential resonance, providing a natural explanation why the MGT show this phenomenon whereas other toad species do not.
On the other hand, the frequency width of the resonance region is apparently large enough to accommodate the modest variations in magnetoreceptive properties expected within the species.




\section{Experimental prospects}

If a MBT is principally capable of detecting GW, how might we be able to harness this amazing feat?
The toads constantly emit their GW imitations -- especially during mating season -- without being induced by a GW event.
Presumedly, {\it during} an event, the toads' central nervous system would react by changing the default call, e.g. by introducting a phase-shift.
Such a phase-shift could be detected simply with a microphone, the continuous recording of a pond full of MBT, and a signal-processing algorithm that follows the trains of croaks looking for phase shifts.

The MP3 audio file used above also contains a brief recording of a pond full of toads.
A sonogramme of this recording is shown in Fig.\,\ref{fig:pond}.
Most of the toad calls can be seen as clearly individual events and no phase-shifts are visible.
Since no GW event was reported during the recording, we can clearly see how powerful the use of toad calls can be for the detection or non-detection of such signals.  

Placing hundreds or thousands of toads in a controlled laboratory environment would undoubtedly improve the detection sensitivity.
Indeed, a facility with millions of toads would probably be able to determine even the direction of the signal, at a tiny fraction of the cost of a set of optical interferometer arrays.

\vfill

\section{Conclusions}

We have shown that Mexican Burrowing Toads produce a croak that highly resembles the gravitational wave signals produced by merging black holes.
There is a plausible biological mechanism that could permit this species to detect gravitational waves via a coherent magnetronic amplification process in neurologically embedded ferromagnetic materials.
That this species alone would show this ability is easily explained as being due to the sensitivity of the resonance conditions to varying ferromagnetic properties.
Although we have ruled out that the MBT are mimicing the {\it exact} waveforms of gravitational events, we nevertheless conclude that these toads have been listening to and mimicing the GW signals of merging black holes and neutron stars for hundreds of thousands if not millions of years.

There are many species that show remarkable sensitivity to ambient magnetic fields but until now it was unthinkable that this sensitivity would extend to gravitational waves.
Further study of these toads is required to determine whether there is some evolutionary advantage accrued by showing this behavior or whether these animals have been given access to the most dramatic events in the cosmos by a remarkable fluke of nature.

If MBT have a sensitivity to gravitational waves, the normal non-induced croaking of these animals should be interrupted by the occurrence of an event, producing an easily detectable phase shift.
We thus encourage the use of these animals as extremely sensitive and unbelievably inexpensive biological gravitational wave detectors.


\section*{Acknowledgements}

We made use of the tools provided by the public domain {\tt astropy}, {\tt emcee}, {\tt matplotlib},  {\tt numpy} and {\tt scipy} python libraries.
The cores of the sonogramme and cepstogramme python scripts were written by {\tt ChatGPT} \footnote{https://openai.com}, one of the major motivations to perform this study.

Please note that the {\bf idea} 
for this paper was {\bf definitely not} obtained from {\tt ChatGPT}. 
We did ask {\tt ChatGPT} {\it afterwards} what it would suggest as good topics (our favourite was ``Temporal Back-Reaction in Citation Counts: Citing Papers Before They're Written''), but we felt ours was already better than the many reasonably good suggestions.


\section*{Data Availability}

The data and python scripts used are available from the authors upon request.

\section*{Ethics Statement}

This study did not involve any actual animal subjects other than those that were passively recorded {\it in situ} during the audio recordings.
No animals were harmed during the research process.
No animal testing was performed in the development of this study.
We always wanted to include this kind of statement in one of our publications and are grateful that we are able to do so now.
\hfill
\newpage



\bibliographystyle{unsrt}

\end{document}